\documentclass[%
reprint,
superscriptaddress,
 amsmath,amssymb,
 aps,
noeprints,
floats,
10pt]{revtex4-2}

\usepackage[utf8]{inputenc}
\usepackage{physics}
\usepackage{amsmath}
\usepackage{caption}
\usepackage{bm}
\usepackage{upgreek}
\usepackage{xr}
\usepackage[dvipsnames]{xcolor}
\usepackage{subcaption}
\usepackage{float}
\usepackage{tablefootnote}
\usepackage{footnote}
\usepackage{graphicx}
\usepackage{braket}
\usepackage[symbol]{footmisc}
\usepackage[export]{adjustbox}
\usepackage{tabularx}
\usepackage{lipsum}  
\newcommand{\comment}[1]{}
\usepackage{multirow}
\usepackage{makecell}
\usepackage{color,soul}

\begin{document}

\title{Non-empirical prediction of the length-dependent ionization potential \\ in molecular chains}

\author{Guy Ohad}
    \affiliation{Department of Molecular Chemistry and Materials Science, Weizmann Institute of Science, Rehovoth 76100, Israel}
\author{Michal Hartstein}
    \affiliation{Department of Molecular Chemistry and Materials Science, Weizmann Institute of Science, Rehovoth 76100, Israel}
\author{Tim Gould}
    \affiliation{Queensland Micro- and Nanotechnology Centre, Griffith University, Nathan, QLD 4111, Australia}
\author{Jeffrey B. Neaton}
    \affiliation{Department of Physics, University of California, Berkeley, Berkeley, California 94720, USA}
    \affiliation{Materials Sciences Division, Lawrence Berkeley National Laboratory, Berkeley, California 94720, USA}
    \affiliation{Kavli Energy NanoSciences Institute at Berkeley, University of California, Berkeley, Berkeley, California 94720, USA}
\author{Leeor Kronik}
   \affiliation{Department of Molecular Chemistry and Materials Science, Weizmann Institute of Science, Rehovoth 76100, Israel}

\begin{abstract}
The ionization potential of molecular chains is well-known to be a tunable nano-scale property that exhibits clear quantum confinement effects. State-of-the-art methods can accurately predict  the ionization potential in the small molecule limit and in the solid-state limit, but for intermediate, nano-sized systems prediction of the evolution of the electronic structure between the two limits is more difficult. Recently, optimal tuning of range-separated hybrid functionals has emerged as a highly accurate method for predicting ionization potentials. This was first achieved for molecules using the ionization potential theorem (IPT) and more recently extended to solid-state systems, based on an \textit{ansatz} that generalizes the IPT to the removal of charge from a localized Wannier function. Here, we study one-dimensional molecular chains of increasing size, from the monomer limit to the infinite polymer limit using this approach. By comparing our results with other localization-based methods and where available with experiment, we demonstrate that Wannier-localization-based optimal tuning is highly accurate in predicting ionization potentials for any chain length, including the nano-scale regime.
\end{abstract}

\maketitle
\thispagestyle{plain}
\pagestyle{plain}

\section{Introduction}

Accurate prediction of the electronic properties of molecular and solid-state systems is of crucial importance in electronics and optoelectronics. In particular, the ionization potential (IP), electron affinity (EA) and therefore the fundamental gap, defined as the difference between the IP and EA, are key quantities in understanding  materials and device properties. \emph{Ab initio} many-body perturbation theory, typically within the $GW$ approximation \cite{hedin1965new, hybertsenFirstPrinciplesTheoryQuasiparticles1985, aulburQuasiparticleCalculationsSolids2000a, onida2002electronic, louie2005quasiparticle,golzeGWCompendiumPractical2019,martinInteractingElectronsTheory2016b}, can be a highly accurate method for predicting these quantities. But owing to its relatively high computational cost, there is an ongoing interest in developing accurate enough approximations within density functional theory (DFT) (see, e.g., Refs.~\cite{kuemmel_kronik_2008,tran_blaha_2009,perdewUnderstandingBandGaps2017,onida2002electronic, miceli_pasquarello_2018, colonna2019koopmans, liLocalizedOrbitalScaling2018, lebeda_kummel_2023right}), that can serve as an inexpensive yet potentially still accurate alternative for computing these important properties. 

Within the framework of DFT, we focus on optimal tuning (OT) \cite{kronik_stein_refaely-abramson_baer_2012} of (screened) \cite{kronik_kuemmel_2018} range-separated hybrid ((S)RSH) functionals, where screening is included for the case of bulk solids. OT-(S)RSH has been shown to be a highly accurate, non-empirical method for predicting electron removal/addition energies and therefore fundamental gaps for a variety of molecular systems (see, e.g., Refs.~\cite{stein_baer_2010, refaely_kronik_2011, kronik_stein_refaely-abramson_baer_2012, autschbach_srebro_2014, phillips_dunietz_2014, foster_allendorf_2014, korzdorfer_bredas_2014, faber2014excited, alipour2017shedding}) and molecular solids (see, e.g., Refs~\cite{refaely-abramson_kronik_2013,luftner_puschnig_2014,kronik2016excited,bhandariFundamentalGapsCondensedPhase2018,kronik_kuemmel_2018, coropceanu2019charge, franco2023unveiling}). Generally, RSH functionals allow for a different combination of exchange and correlation approximations at different ranges of electron-electron interactions and therefore offer flexibility in choosing appropriate functional parameters \cite{toulouse_savin_2004long,vydrov_scuseria_2006importance,chai_headgordon_2008systematic,rohrdanz_herbert_2009long}. OT-(S)RSH allows one to choose such parameters non-empirically by enforcing two conditions: the correct asymptotic behaviour of long-range interactions \cite{levy_sahni_1984, Almbladh_von_Barth_1985, kronik_kummel_2020} and the ionization potential theorem (IPT) \cite{perdew_balduz_1982, levy_sahni_1984, Almbladh_von_Barth_1985, perdew_levy_1997}. The latter has been shown to be particularly crucial not only for IP predictions, but also for accurate EA and therefore gap predictions \cite{stein_baer_2012}. 

Unfortunately, optimal tuning based on straightforward application of the IPT fails in the solid-state limit. This is because, owing to the natural delocalization of electronic orbitals in this limit, the IPT is trivially satisfied for any choice of functional parameters, regardless of the accuracy (or lack thereof) of the obtained electronic structure \cite{mori-sanchez_yang_2008, kraisler_kronik_2014, vlcek_eisenberg_steinle-neumann_baer_2015, gorling_2015}. To overcome this significant limitation, a Wannier-localization-based optimal tuning of SRSH (WOT-SRSH) has been proposed \cite{wing_2021}. This approach enforces a generalized IPT \textit{ansatz} \cite{ma_wang_2016}, based on a constrained removal of charge from a localized Wannier function \cite{marzari_vanderbilt_2012}. WOT-SRSH has recently been shown to be highly successful in predicting band gaps and optical spectra of solids, both alone \cite{wing_2021, ohad_2022_wotsrsh_haps, ohad_gant_2023_metal_oxides} and as an optimal starting point to $GW$ calculations \cite{ohad_gant_2023_metal_oxides, gant2022optimally}, without any empiricism.

The success of OT-RSH in the molecular limit and WOT-SRSH in the solid-state limit immediately raises important  questions as to the utilization of (W)OT approaches for intermediate-size systems. Specifically, for systems of increasing size, at which point does localization become necessary for optimal tuning? How do OT and WOT calculations compare with one another and how well do they predict the evolution of electronic properties with system size, compared to experiment and/or other benchmark calculations? 

A class of systems where such questions arise naturally and can be examined systematically is that of linear oligomers, i.e., linear molecular chains composed of a variable number of repeating units of a given monomer. Indeed, such systems have been previously used to test the accuracy of various approaches within DFT \cite{dafilho2007hole, stein_baer_2010, korzdorfer2011long, bilgicc2011first, dequeirozChargetransferExcitationsLowgap2014, vlcek2016spontaneous, liLocalizedOrbitalScaling2018, su2020preserving, mei2021describing}. 

Here, we use these benchmark systems to answer the above questions by employing OT- and WOT-RSH to compute the IPs of three different one-dimensional molecular chains of increasing length, from the monomer limit to the infinite polymer limit. By comparing our results to other methods and to experiment where available, we show that OT- and WOT-RSH yield essentially identical results for shorter chains, but deviate from each other for larger chains, with WOT-RSH yielding a correct convergence to the infinite polymer limit and providing consistently more accurate results than OT-RSH, as compared to reference theoretical results.

\section{Methods}

\subsection{Benchmark systems}

We study three types of one-dimensional molecular chains: Linear alkanes, \textit{trans}-oligoacetylenes (tOAs) and oligothiophenes (OLTs), the chemical formulas of which are C$_{2n}$H$_{4n+2}$, C$_{2n}$H$_{2n+2}$, and C$_{8n}$H$_{4n+2}$S$_{2n}$, respectively (see inset of Fig.~\ref{fig:ip}). For clarity, throughout we use the term ``polymer'' to refer only to the limit of $n \rightarrow \infty$, namely polyethylene, \textit{trans}-polyacetylene and polythiophene, respectively. See the Supporting Information (SI) \cite{SupportingInfo} for more details.

\subsection{Range-separated hybrid functionals}

In RSH functionals \cite{savin1995density, yanaiNewHybridExchange2004}, the Coulomb operator is partitioned into two terms, typically by exploiting the error function, $\textrm{erf}$, in the form
\begin{equation}
     \frac{1}{r} = \underbrace{\frac{\alpha+\beta\textrm{erf}(\gamma r)}{r}}_{\textrm{xx}} + \underbrace{\frac{1-[\alpha+\beta\textrm{erf}(\gamma r)]}{r}}_{\textrm{SLx}},
     \label{eq:CoulombPartition}
\end{equation}
where $r$ is the interelectronic distance and $\alpha,\beta,\gamma$ are free parameters. While Eq.~\eqref{eq:CoulombPartition} represents a trivial identity, this split of the Coulomb repulsion allows for use of different approximations for the electron exchange associated with each term. For the first term, we use Fock (exact) exchange (xx), whereas for the second we use semilocal exchange (SLx). This leads to two limiting-case fractions of exact exchange: $\alpha$ for short-range (SR) interactions ($r \rightarrow 0$) and $\alpha+\beta$ for long-range (LR) interactions ($r \rightarrow \infty$). These two limits are interpolated smoothly via the error function, with the transition governed by the range-separation parameter, $\gamma$. Accordingly, the exchange energy of RSH is expressed as
\begin{equation}
     E_\textrm{x}^\textrm{RSH} = \alpha E_\textrm{xx}^\textrm{SR} + (1-\alpha) E_\textrm{SLx}^\textrm{SR} + (\alpha + \beta) E_\textrm{xx}^\textrm{LR} + (1-\alpha-\beta) E_\textrm{SLx}^\textrm{LR}.
\end{equation}

In this work, we choose $\alpha$ as 0.25 throughout, as in global and short-range hybrid functionals \cite{perdew_burke_hybrid_1996, adamo_barone_1999, heyd_ernzerhof_2006}, in order to retain a useful balance between exchange and correlation in the short range \cite{refaely2012quasiparticle}. The correct asymptotic limit of the potential is attained by setting $\alpha+\beta=1/\varepsilon$, where $\varepsilon$ is the scalar dielectric constant \cite{refaely-abramson_kronik_2013}. In this work we set $\varepsilon = 1$, the appropriate choice for an isolated molecule in vacuum \cite{levy_sahni_1984, Almbladh_von_Barth_1985, kronik_kummel_2020}, which is correct also for the polymers, where asymptotic screening vanishes owing to the low dimensionality (see, e.g., Ref.\ \cite{qiu2016screening}, for two dimensions). We use the range-separated version \cite{henderson2008generalized, iikura2001long} of the Perdew–Burke–Ernzerhof (PBE) exchange functional \cite{perdew_burke_ernzerhof_1996} to treat the semilocal exchange components in the RSH, along with full PBE semilocal correlation. Non-empirical methods employed to select $\gamma$ are discussed below.

Once the three parameters are selected, one can compute any property of interest. In this work, we focus on the eigenvalue corresponding to the highest occupied molecular orbital (HOMO), the negative of which provides a prediction for the IP in an optimally-tuned functional.

\subsection{Optimal tuning of RSH}

As mentioned in the introduction, in the OT-RSH approach $\gamma$ is selected to enforce the IPT, which is an exact physical condition in DFT. Here, we enforce the IPT for the neutral system, namely we seek $\gamma$ such that 
\begin{equation}
\label{eq:IPT}
     \Delta J^\gamma \equiv E^\gamma(N-1)-E^\gamma(N)+\epsilon_H^\gamma = 0,
\end{equation}
where $E^\gamma(N)$ and $E^\gamma(N-1)$ are the total ground-state energies for the neutral system with $N$ electrons and the singly ionized cation, respectively, and $\epsilon_H^\gamma$ is the HOMO eigenvalue for the $N$-electron system. See the SI \cite{SupportingInfo} for more computational details.

\begin{figure*}[htbp!]
\includegraphics[width=0.99\textwidth]{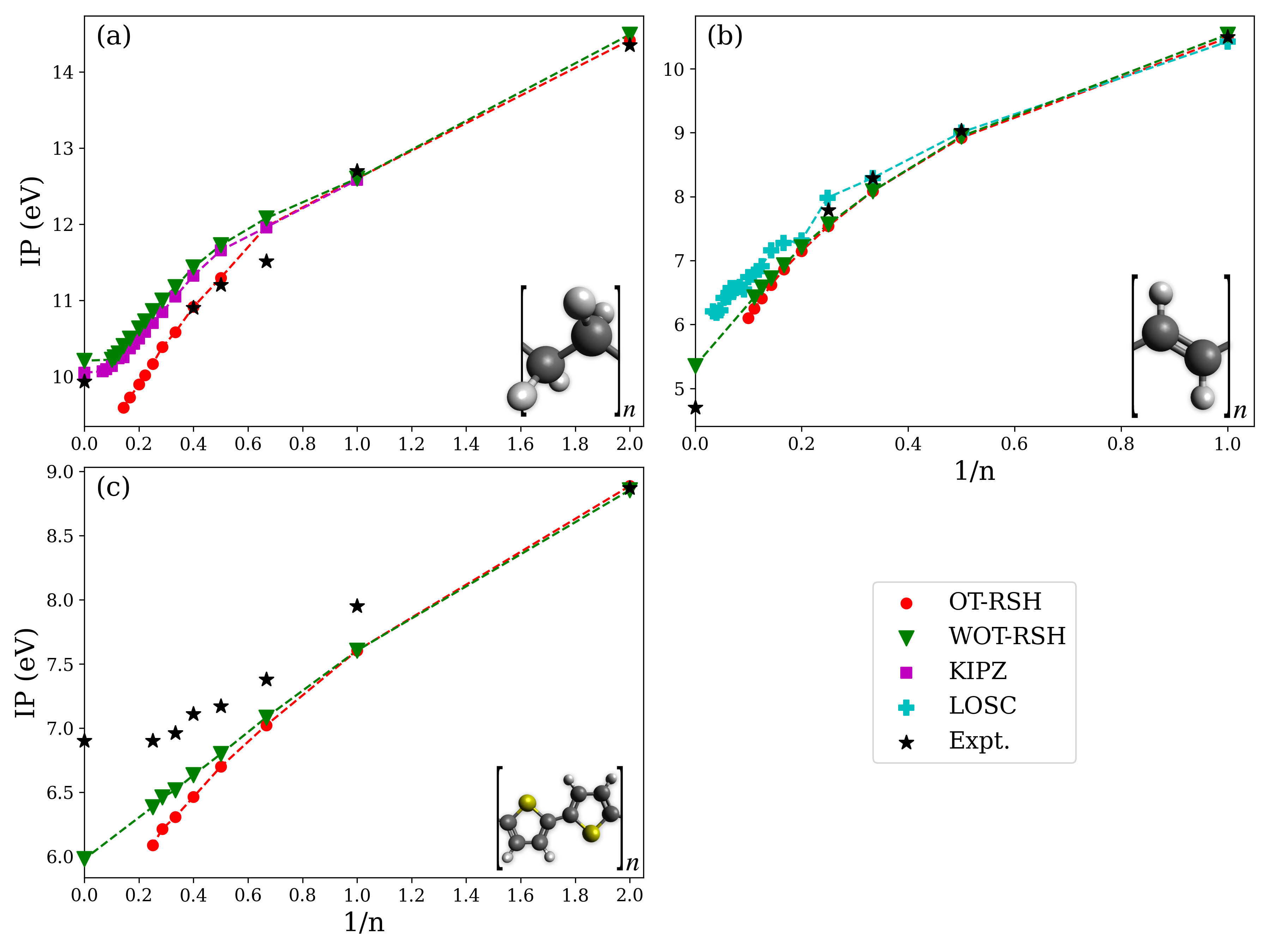}
        \caption{\label{fig:ip} Ionization potential, as a function of the inverse of $n$, the number of repeating units for (a) alkanes, (b) \textit{trans}-oligoacetylenes (tOAs), and (c) oligothiophenes (OLTs).
        Computed results are given by the negative of the HOMO energy based on OT-RSH (red circles), WOT-RSH (green triangles), KIPZ (magenta squares,  from Ref.~\cite{nguyen_marzari_2018}), and LOSC (cyan plus signs, from  Ref.~\cite{liLocalizedOrbitalScaling2018}). Experimental results (black stars) are taken from the following sources: For alkanes: $n=0.5$: Ref.~\cite{potts1972photoelectron}, $n=1$: Ref.~\cite{baker1968electronic}, $n=1.5,2,2.5$: Ref.~\cite{nist}, $n\rightarrow\infty$: Refs.~\cite{partridge1966vacuum, fujihira1972photoemission, seki1986valence}. For tOAs: $n=1,2$: Ref.~\cite{nist}, $n=3$: Ref.~\cite{beez1973ionization}, $n=4$: Ref.~\cite{jones1979study}, $n\rightarrow\infty$: Ref.~\cite{Encyclopedia_polymer_vol5}. For OLTs: Refs.~\cite{dafilho2007hole, jones1990determination}. Inset: schematic view of the repeating unit of each chain, showing carbon atoms in black, hydrogen atoms in grey, and sulfur atoms in yellow.}
\end{figure*}

\subsection{Wannier-localization-based optimal tuning of RSH}

As mentioned above, in the bulk limit the IPT is trivially satisfied for any choice of $\gamma$, which precludes the predictive selection of a unique range-separation value based on Eq.~\eqref{eq:IPT}. Many authors have explored localized orbitals as a means of circumventing this limitation of the IPT, in the context of different approaches within DFT  \cite{anisimov_kozhevnikov_2005, cococcioni_de_gironcoli_2005,ma_wang_2016, weng_wang_2017, liLocalizedOrbitalScaling2018, miceli_pasquarello_2018, nguyen_marzari_2018, bischoff_pasquarello_2019, bischoff_pasquarello_2019_perovskites, elliott2019koopmans, weng_wang_2020, su2020preserving, bischoff2021band, colonna2022koopmans, mahler2022localized, yang2022one, degennaro2022bloch, linscott2023koopmans, wing_2021}. In the context of SRSH functionals, Wing \textit{et al.}~\cite{wing_2021} adopted an \textit{ansatz} \cite{ma_wang_2016} that generalizes the IPT to the removal of an electron from a state that corresponds to a localized Wannier function, namely
\begin{equation}
\label{eq:ansatz}
  \Delta I^\gamma \equiv  E^\gamma[\phi](N-1)- E^\gamma(N)+ \bra{\phi} \hat{H}_{N}^\gamma \ket{\phi} = 0,
\end{equation}
where $\phi$ is the Wannier function for which the expectation energy with respect to the DFT Hamiltonian of the $N$-electron system, $\bra{\phi} \hat{H}_{N}^\gamma \ket{\phi}$, is the largest. $E^\gamma[\phi](N-1)$ is the total energy of the \emph{constrained} $N-1$-electron system, which differs from the ground-state $N-1$-electron system in that an electron has been removed from the Wannier function \cite{wing_2021}. This constraint is imposed via a Lagrange multiplier that controls the occupancy of the Wannier function, which is constructed of the occupied-orbital manifold using the PBE functional. Additional computational details are given in the SI \cite{SupportingInfo}.

\section{Results and Discussion}

\begin{figure*}[htbp!]
\includegraphics[width=0.99\textwidth]{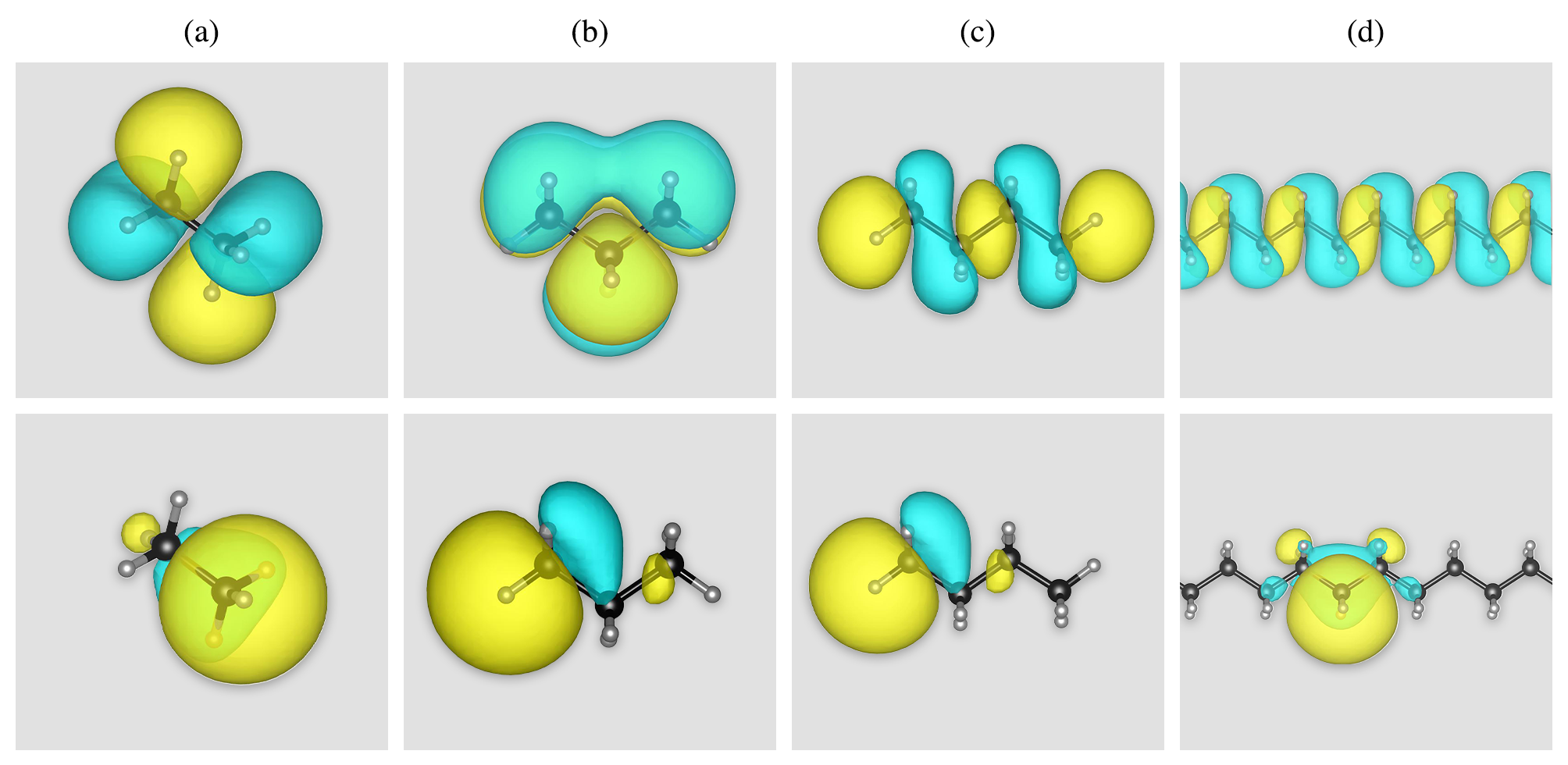}
        \caption{\label{fig:homo_wannier_alkanes} HOMO (top row) and highest expectation energy Wannier function (bottom row) for selected alkanes: (a) ethane ($n=1$), (b) propane ($n=1.5$), (c) butane ($n=2$), and (d) polyethylene ($n \rightarrow \infty$). Carbon and hydrogen atoms are shown in black and grey, respectively. The wavefunction isosurface is shown in light blue and yellow for a value of 2.0.}
\end{figure*}

Fig.~\ref{fig:ip} shows the computed IP, based on the negative of both the OT-RSH and WOT-RSH HOMO energy, as a function of the inverse of $n$, the number of repeating units, for each of the three systems studied in this work. These results are compared with those obtained from two other localization-based methods: the integer Koopmans plus Perdew-Zunger correction (KIPZ) method for the alkanes, taken from Ref.~\cite{nguyen_marzari_2018}, and the localized orbital scaling correction (LOSC) approach for the tOAs, taken from Ref.~\cite{liLocalizedOrbitalScaling2018}. The computational results are further compared with experiment where available.

First, we compare the OT-RSH and WOT-RSH results. We observe two trends that are common to all three systems. First, the two methods predict essentially identical IPs for the shorter chains, where the HOMOs are ``naturally'' localized owing to the small system size. This is a significant observation, because it demonstrates the validity and generality of the IPT \textit{ansatz} used in WOT-RSH, even in a realm for which it was not designed and in which it is not strictly necessary. Second, the deviation between the two methods increases with chain length and becomes as large as $\sim$0.8 eV for the case of alkanes with $n=7$. We attribute this to the delocalization of the HOMO in the longer chains, shown in Fig.~\ref{fig:homo_wannier_alkanes} for selected alkanes and in the SI for selected tOAs and OLTs. This delocalization ultimately prevents the use of OT-RSH altogether for large enough chains, as the IPT is approaching the point where it is trivially obeyed and a numerically stable determination of the range-separation parameter is no longer possible. In contrast, the WOT-RSH relies on a Wannier function that is localized by construction and changes little with $n$, as also shown in Fig.~\ref{fig:homo_wannier_alkanes} for selected alkanes and in the SI for selected tOAs and OLTs. As a result, the WOT-RSH procedure is numerically stable and physically meaningful for any $n$, including $n \rightarrow \infty$, i.e., the polymer limit.

Interestingly, while the trend of OT- and WOT-RSH results deviating from one another is common to the three systems, it appears to be  more abrupt for the alkanes, where the deviation starts already at relatively short chains, but more gradual and at larger $n$ for the two other systems. We note that the abrupt deviation in the alkanes occurs between $n=1.5$ and $n=2$. This can be associated with an abrupt change in the symmetry of the HOMO, owing to orbital reordering, between these two chains, as demonstrated in Fig.~\ref{fig:homo_wannier_alkanes} (b) and (c). The same symmetry as in $n=2$ is then maintained for all $n > 2$. The symmetry of the HOMO for the tOAs and OLTs, on the other hand, is unchanged for all $n$, as demonstrated in the SI.

Next, we compare our results to benchmark computational data. As shown in Fig.~\ref{fig:ip}, the general trend of the IP saturating with increasing chain length, up to the polymer limit, is clearly captured. This trend is less observed in the tOAs, in agreement with the results of Ref.~\cite{vlcek2016spontaneous}, which showed that the saturation occurs in longer chain lengths that are outside the range studied in this work. Furthermore, the WOT-RSH results agree very well quantitatively with previous localization-based schemes - to $\sim$0.1 eV with KIPZ results and $\sim$0.2 eV with LOSC, on average. Conversely, and as expected based on the above discussion, the OT-RSH results do not extrapolate to the correct polymer limit.

Finally, we compare the computational results to experimental ones. For alkanes, the experimental results agree well with all computational methods for $n=0.5$ and $n=1$. For $n=1.5$, all computational methods appear to agree, but predict a value larger than experiment by $\sim$0.5 eV. For $n=2$ and $n=2.5$, WOT-RSH and KIPZ overestimate experiment by a similar amount, while OT-RSH is in better agreement with it. This, however, may be accidental, given the fact that the $n=1.5$ result of OT-RSH overestimates experiment. The IP for $n \rightarrow \infty$, polyethylene, agrees well with several experimental estimations \cite{partridge1966vacuum, fujihira1972photoemission, seki1986valence}.

For the tOAs, agreement with existing experimental values for finite chains is consistently good for all computational methods. The experimental value for \textit{trans}-polyacetylene is taken from solid-state measurements, where the IP can be smaller by hundreds of meV from the gas-phase IP \cite{Encyclopedia_polymer_vol5}, possibly explaining the deviation from the WOT-RSH prediction. For the OLTs, with the exception of the $n=0.5$ monomer, our results underestimate experimental ones by more than 0.3 eV. Whether this discrepancy is related to structural differences, thermal effects, experimental uncertainties, or theoretical limitations is at present unknown. Even so, the WOT-RSH results agree qualitatively and semi-quantitatively with the experimental ones, whereas the OT-RSH results do not. Overall, then, the WOT-RSH results provide good agreement with experimental trends, where available, throughout.

\section{Conclusions}

We have compared two optimal tuning methods of RSH, namely OT-RSH and WOT-RSH, for the computation of the IP of one-dimensional molecular chains of increasing length. We have demonstrated the known failure of OT-RSH for long chains, owing to orbital delocalization. We found, however, that WOT-RSH is successful in predicting accurate IPs throughout the evolution of the chain length, not only in the polymer limit for which it was originally designed, but throughout the entire range of oligomers, from monomer to polymer. Specifically, WOT-RSH results agree with both experimental trends and past localization-based computational schemes. This provides a first step in the application of optimal tuning to nano-sized objects where neither the molecular nor the bulk limits apply.

\section*{Acknowledgments}
This work was supported via U.S.-Israel NSF–Binational Science Foundation Grant No. DMR-2015991 and by the Israel Science Foundation. The authors thank Nadav Ohad for graphic design. T.G. and L.K. were supported by an Australian Research Council (ARC) Discovery Project (DP200100033). T.G. was supported by an ARC Future Fellowship (FT210100663). L.K. was additionally supported by the Aryeh and Mintzi Katzman Professorial Chair and the Helen and Martin Kimmel Award for Innovative Investigation.

\bibliography{references.bib}
\end{document}